\pdfoutput=1
\documentclass[prb,amssymb,reprint]{revtex4-1}

\usepackage{graphics}
\usepackage{mathtools}
\usepackage{float}
\usepackage{amsmath}

\begin{document}
\title{Anisotropic, multi-carrier transport at the (111) LaAlO$_3$/SrTiO$_3$ interface}

\author{S. Davis}
\email[]{samueldavis2016@u.northwestern.edu}
\author{V. Chandrasekhar}
\email[]{v-chandrasekhar@northwestern.edu}
\affiliation{Graduate Program in Applied Physics and Department of Physics and Astronomy, Northwestern University, 2145 Sheridan Road, Evanston, IL 60208, USA}
\author{Z. Huang}
\author{K. Han}
\author{Ariando}
\affiliation{NUSNNI-Nanocore, National University of Singapore 117411, Singapore}  
\affiliation{Department of Physics, National University of Singapore 117551, Singapore } 
\author{T. Venkatesan}
\affiliation{NUSNNI-NanoCore, National University of Singapore 117411, Singapore}
\affiliation{Department of Physics, National University of Singapore 117542, Singapore}
\affiliation{Department of Electrical and Computer Engineering, National University of Singapore 117576, Singapore}\affiliation{Department of Material Science and Engineering, National University of Singapore 117575, Singapore} 
\affiliation{NUS Graduate School for Integrative Sciences \& Engineering, National University of Singapore 117456, Singapore}

\begin{abstract}
	The conducting gas that forms at the interface between LaAlO$_3$ and SrTiO$_3$ has proven to be a fertile playground for a wide variety of physical phenomena. The bulk of previous research has focused on the (001) and (110) crystal orientations.  Here we report detailed measurements of the low-temperature electrical properties of (111) LAO/STO interface samples.  We find that the low-temperature electrical transport properties are highly anisotropic, in that they differ significantly along two mutually orthogonal crystal orientations at the interface.  While anisotropy in the resistivity has been reported in some (001) samples and in (110) samples, the anisotropy in the (111) samples reported here is much stronger, and also manifests itself in the Hall coefficient as well as the capacitance.  In addition, the anisotropy is not present at room temperature and at liquid nitrogen temperatures, but only at liquid helium temperatures and below. The anisotropy is accentuated by exposure to ultraviolet light, which disproportionately affects transport along one surface crystal direction.  Furthermore, analysis of the low-temperature Hall coefficient and the capacitance as a function of back gate voltage indicates that in addition to electrons, holes contribute to the electrical transport. 
\end{abstract}

\date{\today}%
\maketitle

\section{Introduction}
The two-dimensional electron gas that forms at the interface of the band insulators LaAlO$_3$ (LAO) and SrTiO$_3$ (STO)\cite{Ohtomo,Thiel} has attracted a great deal of attention due to the rich landscape of competing physical phenomena that have been observed, including superconductivity\cite{Reyren,Caviglia}, magnetism\cite{Ariando,Pav,Baner,Joshua}, tunable spin-orbit interaction\cite{Gop}, gate-tunable superconductor-to-insulator transistions\cite{Schne,Dikin}, and the coexistence of ferromagnetism and superconductivity\cite{Li,Bert,Mehta}.   

Most studies so far have been on (001) oriented LAO/STO, focusing both on the origins of the 2DEG\cite{ liu,pent,Joshua2,Dagan,Kala,chen} as well as the aforementioned phenomena.  However, (110) and (111) LAO/STO interfaces, which have different symmetries and consequently band structure, are just beginning to be explored,\cite{Herr,Annadi,Gop} but detailed studies of (110) and (111) LAO/STO interfaces as a function of gate voltage have not been reported. There are significant differences between the (001), (110) and (111) oriented LAO/STO interfaces that raise the possibility of interesting new behavior.  Specifically, the symmetries of the band structure at each interface are different. In particular, the orbitals at the (111) interface have hexagonal symmetry (see Fig. \ref{Fig1}(a)),\cite{Rodel,Walker}, with corresponding potentially novel properties.\cite{Doe} Thus it is of interest to study the properties of the (111) LAO/STO interface in detail.

From the point of view of crystal structure and sample morphology, it is not clear whether the properties of LAO/STO structures of any crystal orientation should be isotropic or anisotropic, as there are strong arguments for both.  For (111) LAO/STO interfaces, for example, there is a much larger overlap of the Ti orbitals along the $[\bar{1}\bar{1}2]$ direction than along the $[1\bar{1}0]$ direction (see Fig. \ref{Fig1}(a)).\cite{Walker,Rodel}  Consequently, it might be natural to expect anisotropy in the transport properties between the two directions, such as in the electrical conductance or perhaps even the Hall coefficient, if multiple carrier bands are involved.  However, for crystals with cubic symmetry, such as bulk STO at room temperature, the conductivity should be isotropic.\cite{ashcroft}  On the other hand, bulk STO has a structural transition to a tetragonal phase at $\sim$105 K;\cite{loetzsch} in thin films, or in heterostructures with other materials (e.g., LAO), the transition at this temperature may be to a crystal structure with lower symmetry.  When the symmetry of the crystal structure is lowered, anisotropy in the properties is not surprising.  Finally, for highly disordered samples, as is the case for most of the LAO/STO samples studied so far, any anisotropy might be expected to be averaged out due to scattering between electronic bands.    

Anisotropic transport behavior has been reported in (110) LAO/STO heterostructures,\cite{Gop} but not so far in (111) interface devices.  We report here reproducible anisotropic behavior in the resistance and the Hall coefficient of (111) LAO/STO interface samples that only manifests itself at low temperatures ($\sim$4 K and below), suggesting that it is not associated with the structural transition that occurs at $\sim$105 K.  In addition, we observe anisotropy in the capacitance, a quantity related to the density of states, and hence by definition a quantity that should be isotropic.  The anisotropy in the properties (electrical conductance, Hall coefficient, capacitance) is particularly pronounced below a certain threshold in the back gate voltage $V_g$ applied to the substrate.  This suggests different band edges along the two directions at low temperatures, in contrast to the result from band structure calculations and and ARPES measurements applicable at higher temperatures.

\section{Sample fabrication and measurement}
The 20 monolayer (ML) (111) LAO/STO interface samples used in this study were prepared by pulsed laser deposition using a KrF laser ($\lambda$ = 248 nm).  A LAO single crystal target was used for deposition.  During the deposition, the laser repetition was kept at 1 Hz, laser fluence at 1.8 J/cm$^2$, growth temperature at 650 C, and oxygen pressure at 1 mTorr. The deposition was monitored by in-situ reflection high energy electron diffraction (RHEED).
\begin{figure}[h!]
\center{\includegraphics[width=6cm]{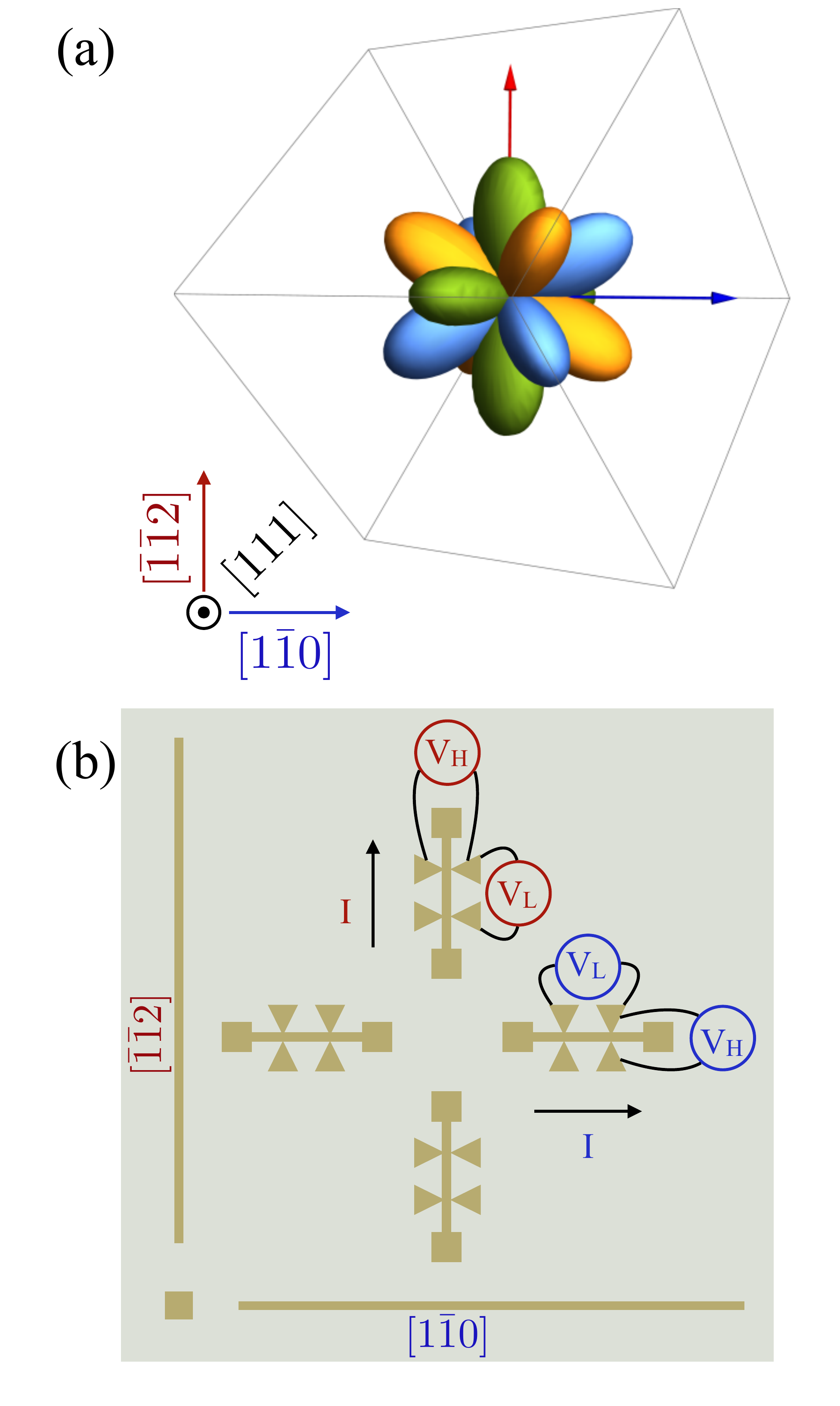}}
\caption{(a)  Schematic representation of the Ti $d_{xy}$, $d_{yz}$ and $d_{xz}$ orbitals in a cubic crystal viewed along the [111] direction.  (b)  Schematic representation of the surface of the (111) LAO/STO sample chip, showing the four Hall bars.  The electrical connections for longitudinal and Hall resistance measurements used for the two Hall bars that are discussed in detail in the text are as marked.  Also shown are the crystal directions determined from x-ray measurements.}
\label{Fig1} 
\end{figure}
For the transport measurements, Hall bars were patterned using photolithography that defined an etch mask used for subsequent argon ion milling.  The argon ion milling etched the unprotected areas to the bare STO, leaving the LAO on top of the Hall bars.  A subsequent photolithography step deposited a Au film on the electrical contacts to enable visual location of the contacts for wire-bonding.  Au was also deposited on the bare etched STO, which enabled us to measure the geometric capacitance as described below, and also to confirm that the bare STO itself did not become conducting after the ion milling step. Six different sample chips were measured and showed similar behavior; here we focus on measurements on one 5x5 mm$^2$ chip on which four Hall bars were fabricated, as shown in Fig. \ref{Fig1}(b).  The substrates were cut so that one edge was along the [1$\bar{1}$0] crystal direction, and the mutually orthogonal edge was along the [$\bar{1}\bar{1}2$] crystal direction; thus for two of the Hall bars the measurement currents were along the [1$\bar{1}$0] direction, while for the other the measurement currents were along the [$\bar{1}\bar{1}2$] direction.  X-ray measurements were used to determine the actual crystal orientation, which is marked in Fig. \ref{Fig1}(b).  While all four Hall bars were measured, Hall bars oriented along the same crystal direction had essentially identical measured properties.  Consequently, we show here data for only the two Hall bars whose electrical contacts are marked in Fig. \ref{Fig1}(b). Figure \ref{SRT}(a) shows an AFM image of the surface of the LAO/STO heterostructure. The Hall bars were fabricated so that the current direction was oriented at 45$^{\circ}\pm 0.5^{\circ}$ to the atomic terraces shown in Fig \ref{SRT}(a). 

The samples were measured in three different cryostats:  a home-built liquid helium cryogenic insert, a Kelvinox MX100 dilution refrigerator, and a Kelvinox 300 dilution refrigerator, although apart from the data in Fig. \ref{SRT}(b), the results reported here are only at liquid helium temperatures.  All measurement rigs were equipped with superconducting solenoids that permitted measurement of the Hall effect.  To measure the electrical transport properties of the Hall bars, an ac current of amplitude $\sim$20 nA was introduced using a home-built low-noise current source with output impedance $>10^{12}$ $\Omega$, and the resulting longitudinal or transverse voltage was detected with a lock-in amplifier after being amplified by a factor of 10 or 100 by a home-built low noise voltage preamplifier.  Due to the very high resistances of the devices, particularly at low temperatures and negative $V_g$, the ac excitation frequency was kept in the range of 1-2 Hz.  The samples were mounted using silver paste on an electrically isolated OFHC copper puck to which a voltage could be applied to gate the interface.  It should be noted that all of the chips show the same qualitative behavior, and measurement of samples on different chips were quantitatively within 10\% of each other.

\begin{figure}[t]
\center{\includegraphics[width=7cm]{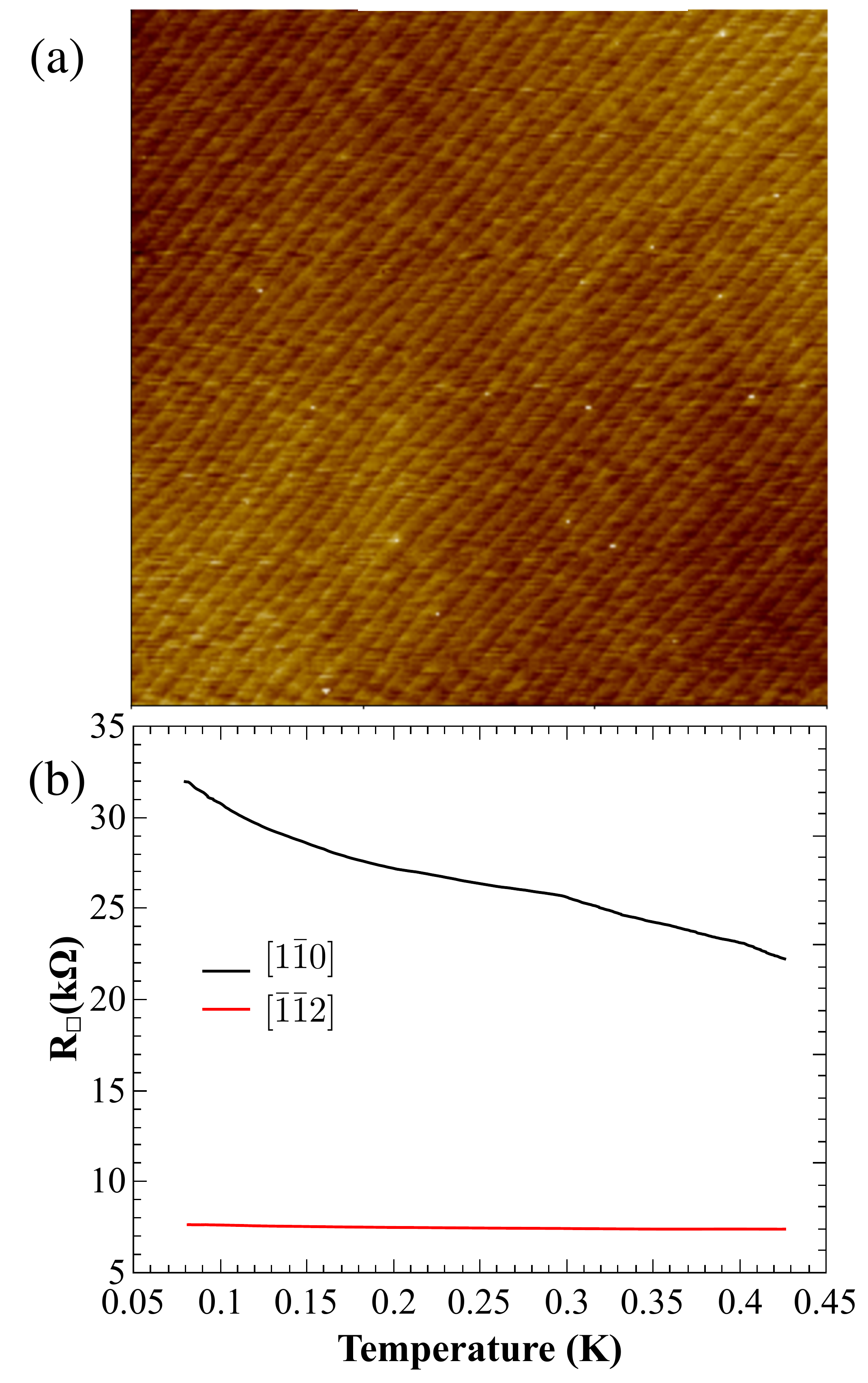}}
\caption{(a)  6$\mu$m X 6$\mu$m AFM topography scan of the LAO/STO sample surface. The $x$ axis corresponds to the [1$\bar{1}$0] direction and the $y$ axis to the [$\bar{1}\bar{1}2$] direction.   (b)  Low temperature resistance vs. temperature sweeps for the two crystal directions.}
\label{SRT} 
\end{figure}

In addition to the longitudinal and Hall resistances, we also measured the $V_g$ dependence of the capacitance $C$ between the back gate and each individual Hall bar.  In order to measure $C$, we used a technique similar to one that has been previously used by other groups to measure the capacitance of GaAs/AlGaAs two-dimensional electron gases \cite{Haigh,Eisen,Li2,Kopp}.  A 100 mV ac voltage was superposed on the dc back-gate voltage $V_g$ (see Fig. \ref{CapRef}(a)) by summing a dc voltage and an ac voltage from the oscillator of an EG\&G digital lock-in amplifier with a home-made summing amplifier, the output of which was used to drive a Kepco 100/1 power supply/amplifier.  The output of the Kepco was then applied to the back-gate. For the other electrode, one contact of the Hall bar was connected to the input of a home-made current preamplifier whose output was fed to the input of the lock-in amplifier.  It was important to ensure that all other contacts of the Hall bar as well as all other Hall bars on the same chip remained open, so that the only current that could flow was to the current preamplifier.  Signals that were in phase and in quadrature with the ac voltage applied to the back gate were then monitored as a function of $V_g$.  The Kepco has a reasonably flat frequency response up until approximately 2.5 kHz; for the greatest sensitivity, we desired to operate at the highest frequency possible, hence the capacitance data for this work was taken at a frequency of 2.3 kHz.  

We tested this measurement configuration on standard capacitors attached to our cryogenic inserts with values close to the measured capacitances of our sample, and the measured values agree within less than 1 \% with their values measured on the bench with commercial capacitance bridges.  In addition to ensuring that our measurement setup was accurate, this also indicated that the capacitance of the wires going down the cryogenic inserts do not significantly alter the measured capacitance.   The detailed analysis in Appendix \ref{CapacitanceAppendix} shows that the component of the ac current in quadrature to the applied ac voltage is a direct measure of $C$, and that the measured capacitance is not significantly affected by the resistance of the Hall bar.

\section{Experimental Results}

\subsection{Longitudinal resistance}
\label{LongRes}
\begin{figure}[h!]
\includegraphics[width=7.5cm]{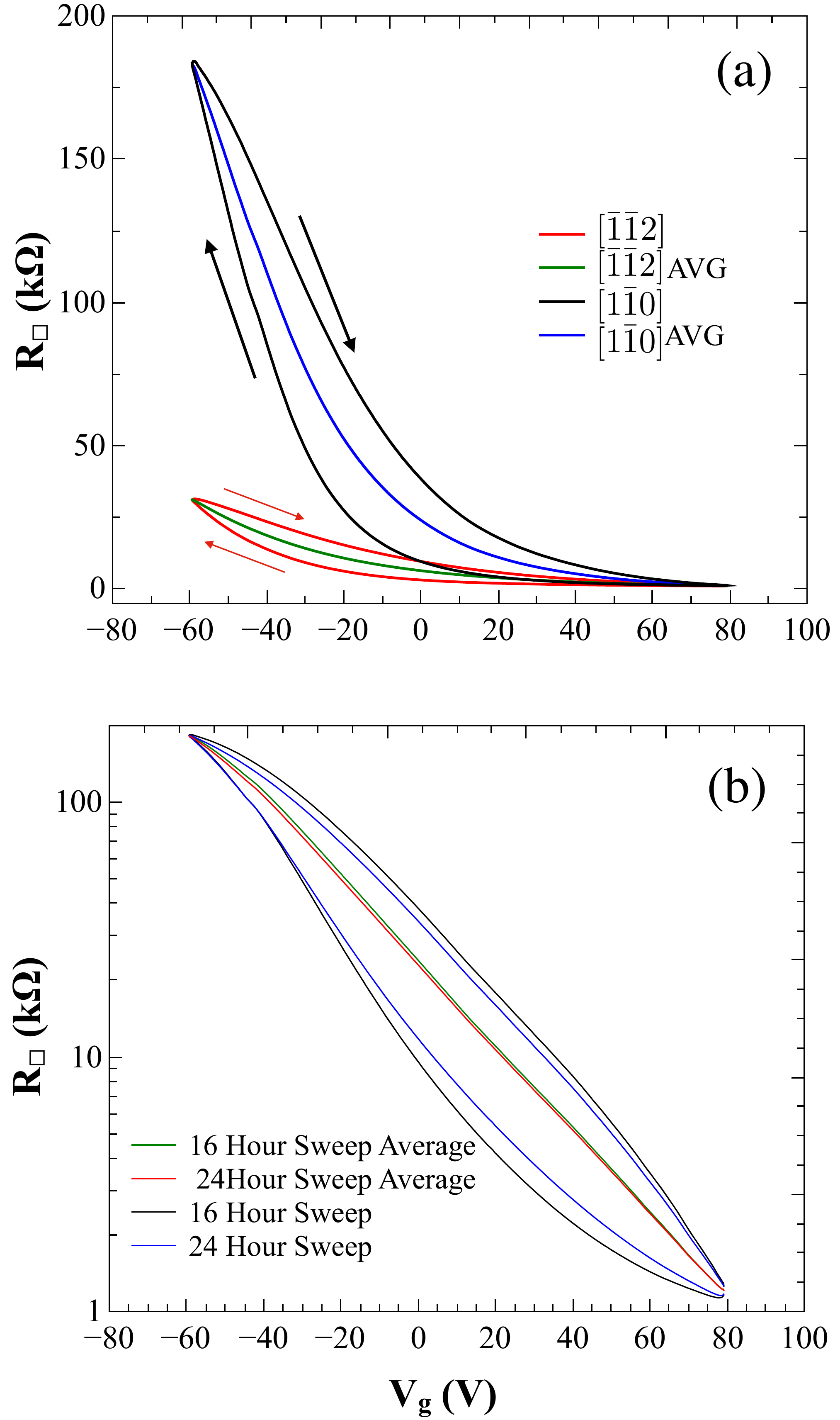}
\caption{\small (a)  Longitudinal sheet resistance $R_{\square}$ of the [$1\bar{1}0$] and [$\bar{1}\bar{1}$2] direction Hall bars as a function of the back gate voltage $V_g$.  The sweep direction is marked by arrows.  The green and blue curves show the average of the up and down sweeps for the two samples.  (b) Logarithmic plot of $R_\square$ vs $V_g$ at two different sweep rates for the $[1\bar{1}0]$ sample. The black/(blue) curve corresponds to a 16/(24) hour long sweep of the back gate with its averaged curve in green/(red). Data taken at 4.4 K.}
\vspace{-0.5cm}
\label{gateh}
\end{figure} 
We start by discussing the gate dependence of the longitudinal resistance.   Figure \ref{gateh}(a) shows the longitudinal sheet resistance $R_\Box$ of the two Hall bars shown in Fig. \ref{Fig1}(b) at 4.4 K as a function of $V_g$.  The first thing to note is that, similar to the (001) samples, $R_\Box$ vs. $V_g$ is strongly hysteretic:  the resistance on sweeping up in $V_g$ is higher than the resistance on sweeping down in $V_g$; up-sweeps reproduced up-sweeps and down-sweeps reproduce down-sweeps if the sweep range and sweep rate remain the same.  We find similar behavior in any measurement that involves sweeping the back gate voltage, in particular, the capacitance measurements discussed later.  In addition, we find that if the gate voltage sweep is stopped at a particular value of $V_g$, $R_\Box$ relaxes to a value corresponding to the average of the up- and down-sweep traces at that value of $V_g$ for that specific sweep rate.  At millikelvin temperatures, this relaxation can take place over hours or even days, particularly at large negative $V_g$.  We speculate that this glassy behavior is due to states that relax slowly to lower energy configurations over very long time scales.  This is rather surprising, as one might expect such processes to be frozen out at millikelvin temperatures.  We plan to return to this interesting issue later.  For the present study, we take advantage of our experimental observation that as the sweep rate is slowed, the hysteresis loops narrow, but the average of the up- and down-sweeps remain the same.  This is shown in Fig.\ref{gateh}(b), which shows two different $R_H$ vs. $V_g$ hysteresis loops taken over 16 hours and 24 hours, and the respective averages over the up and down sweeps, which are essentially identical.  Thus, the averaged curves are measures of the long-term behavior of the devices.  Similar behavior is observed for the gate dependent capacitance measurements.  Consequently, in what follows, we shall present averaged curves for the $V_g$ dependence of the longitudinal resistance (as shown in Fig. \ref{gateh}(a)) and the capacitance.  The Hall data are taken at specific values of $V_g$ after the sample has relaxed, so such averaging is not required.

Returning now to the longitudinal resistance as a function of $V_g$ shown in Fig \ref{gateh}(a), the most striking aspect of the data is that the resistance is highly anisotropic:  while the resistances of both Hall bars are roughly the same at large positive $V_g$, they start to diverge at small positive $V_g$, and differ by more than a factor of 6 at $V_g=-60$ V.  This difference is even larger at millikelvin temperatures, with the resistance in the [$\bar{1}\bar{1}2$] direction changing little, while the resistance in the [$1\bar{1}0$] direction increasing substantially with decreasing temperature, as seen in Fig \ref{SRT}(b).  A longitudinal resistance that depends on crystalline direction has been seen earlier in both (001) and (110) LAO/STO interface samples.  Brinks \textit{et al.} found a strong directional dependence of the resistivity on (001) LAO/STO interfaces grown on LSAT that was associated with step edges in the underlying LSAT substrate\cite{Brinks} the resistance measured along the step edges was significantly larger than the resistance measured perpendicular to the step edges. The difference was largest at low temperatures, but still appreciable at room temperature.  Frenkel \textit{et al.} observed a weak anisotropy\cite{Frenkel} associated with the channeling of currents along microstructural channels that changed on warming above the cubic to orthorhombic transition that occurs in STO at $\sim$ 105 K.  Gopinadhan \textit{et al.} observed a weak anisotropy in the mobility of (110) samples at low temperatures\cite{Gop}.  In contrast to these previous observations, the anisotropy in the electrical characteristics of our (111) samples does not exist at room temperature or liquid nitrogen temperatures, and is reproducible even after repeated warming to room temperature.

\subsection{Hall coefficient}
\label{hall}  
Further evidence of anisotropic behavior can be found in the Hall coefficient $R_H$, which we determine by measuring the Hall resistance in a magnetic field $H$ applied perpendicular to the plane of the sample as a function of $V_g$.   In the range $\mu_0 H \sim \pm  400$ mT, the measured Hall resistance is linear.  We define the Hall coefficient $R_H$ as the slope of the Hall resistance vs. $H$.  
\begin{figure}[h!]
\center{\includegraphics[width=7cm]{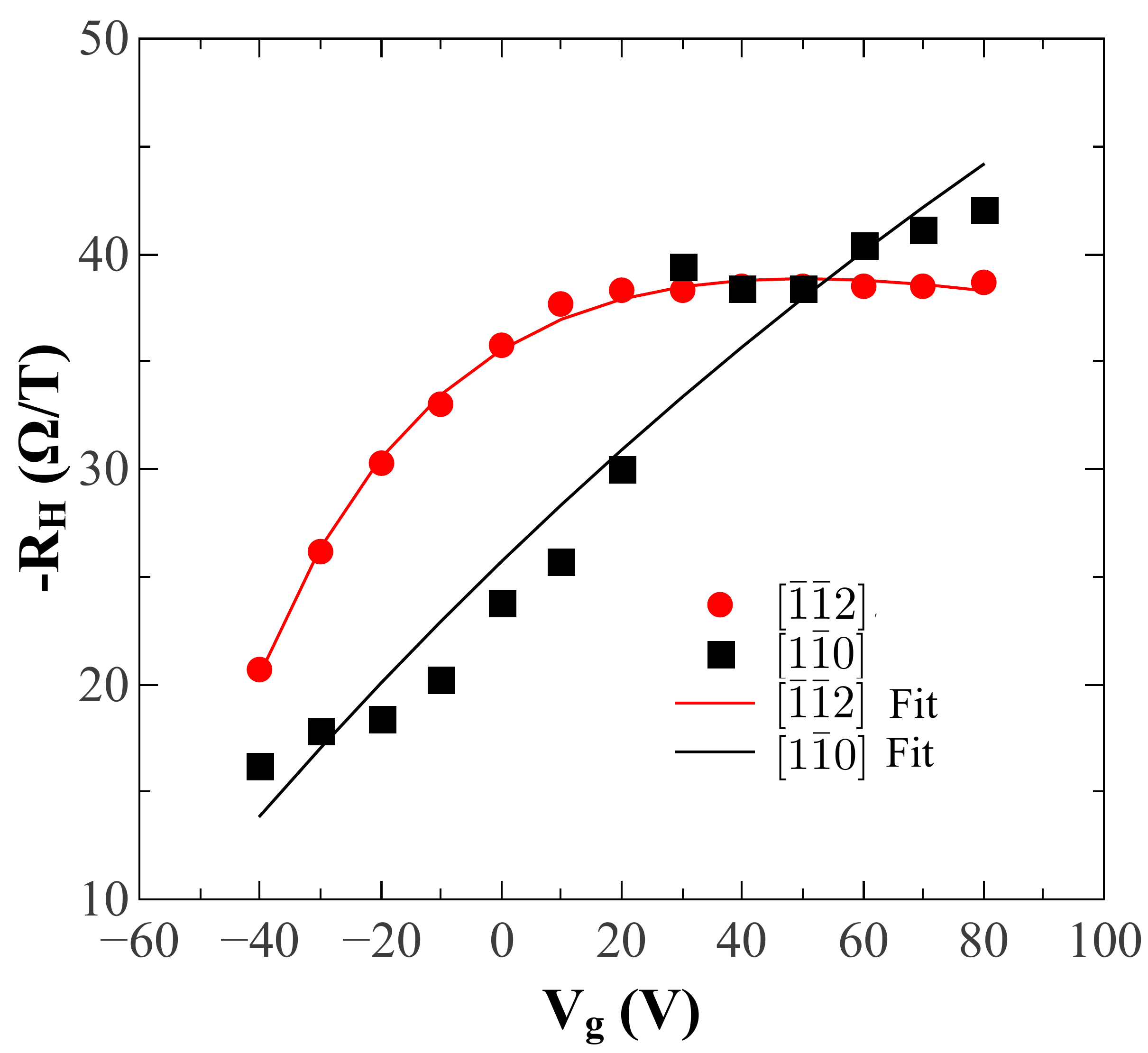}}
\caption{Symbols, measured Hall coefficient $R_H$ of $V_g$ for the two crystal orientations; solid lines are fits to the two-band model described in the text.  All measurements were performed at 4.4 K.}
\label{HallPreUV}
\end{figure}

The solid symbols in Fig. \ref{HallPreUV} show $R_H$ as a function of $V_g$ for both Hall bars.  The magnitude and overall trend of $R_H$ is the same for both crystal orientations, but the detailed dependence on $V_g$ is different:  while $|R_H|$ for the [$\bar{1}\bar{1}2$] direction saturates at large positive $V_g$ and decreases rapidly at negative $V_g$, $|R_H|$ for the [$1\bar{1}0$] direction shows a roughly linear dependence, with a reproducible jump in $|R_H|$  between $V_g=20$ V and $V_g=30$ V. The overall trend is similar to what was observed by Dikin \textit{et al.} in (001) interface samples, but is in contrast with the work done on (001) by Joshua et al. \cite{Joshua2,Dagan}, who found that $R_H$ increases both at high positive $V_g$ as well as large negative $V_g$.

For a single band of charge carriers, $R_H = 1/nq$, where $n$ is the areal density of the carriers, and the sign of $R_H$ is indicative of the sign of the carrier charge.  The sign of $R_H$ in our samples at all $V_g$ is electron-like, but the dependence of $R_H$ on $V_g$ is not as expected for a single band of electrons.  As $V_g$ is increased, one expects that more electrons would be drawn to the interface, so that the electron density $n_e$ would increase.  The increase in $n_e$ should lead to a decrease in the longitudinal resistance, as we observe, but it should also lead to a decrease in $|R_H|\sim 1/n_e$ with increasing $V_g$.  Instead, an \emph{increase} in $|R_H|$ with $V_g$ is observed for both Hall bars.  We note that in earlier experiments on (001) LAO/STO samples, similar anomalous (and in some cases nonmonotonic) dependences of $R_H$ on $V_g$ \cite{Joshua2,Dagan}  was ascribed to multiple electron bands with different mobilities.  Although we agree that the presence of multiple bands is required to explain the anomalous dependence, we cannot see how one can fit the $V_g$ dependence shown in Fig. \ref{HallPreUV} with only \textit{electron} bands, without making the unphysical assumption that the occupation of at least one of the electron bands decreases with increasing $V_g$.  In addition, as discussed below, $R_H$ almost vanishes at large $V_g$ after ultraviolet irradiation, suggesting a compensated system.  Thus we are lead to the conclusion that in addition to any electron bands, at least one hole band must contribute to the electrical transport properties of our samples.

To model the qualitative $V_g$ dependence of $R_H$, we fit the data to the simplest two-band model with one electron band and one hole band:
\begin{equation}
R_H = \frac{R_e \sigma_e^2 + R_h \sigma_h^2}{(\sigma_e + \sigma_h)^2},
\label{eq1}
\end{equation} 
where $R_{(e,h)} = \mp 1/(n_{(e,h)} e)$ are the Hall coefficients and $\sigma_{(e,h)}=n_{(e,h)} e \mu_{(e,h)}$ are the conductivities of the electrons and holes respectively.  Here $n_{(e,h)}$ and $\mu_{(e,h)}$ are the corresponding areal densities and mobilities.  To analyze the gate dependence of the Hall coefficients shown in Fig. \ref{HallPreUV} and \ref{Fig3}  we assume that the electron and hole densities $n_e$ and $n_h$ have a linear dependence on $V_g$. We also make the simplifying assumption that the mobilities of the electrons and holes $\mu_e$ and $\mu_h$ are independent of $V_g$. Thus, there are nominally 6 parameters that can be varied to obtain a fit of $R_H$ as a function of $V_g$: the values of $n_e$ and $n_h$ at $V_g$ = 0, the assumed linear slopes $dn_e/dV_g$ and $dn_h/dV_g$, and the mobilities $\mu_e$ and $\mu_h$. We use the measured value of the 2D conductivity $\sigma = 1/R_\Box$  as a function of $V_g$ to reduce the number of parameters by one, since $\sigma_e$ + $\sigma_h$ = $\sigma$, using as a 5th fitting parameter the ratio of mobilities $\mu_r$ = $\mu_e$/$\mu_h$, subject to the constraint that $n_e$($V_g)e\mu_e$ +$n_h(V_g)e\mu_h$ = $\sigma (V_g)$. The parameters obtained from the fits are given in Table \ref{Table1}.
\begin{table*}[!]
%\begin{ruledtabular*}
\centering
\caption{Pre-UV irradiation Hall fit parameters}
\label{Table1}
\begin{tabular}{|l|r|r|r|r|r|r|}
\hline
 & \multicolumn{1}{c|}{$n_e$ ($V_g=0$) (/cm$^2$)} & \multicolumn{1}{c|}{$n_h$ ($V_g=0$) (/cm$^2$)} & \multicolumn{1}{c|}{$\mu_e$ (cm$^2$/V-s)} & \multicolumn{1}{c|}{$\mu_h$ (cm$^2$/V-s)} & \multicolumn{1}{c|}{$dn_e/dV_g$ (/cm$^2$-V)} & \multicolumn{1}{c|}{$dn_h/dV_g$ (/cm$^2$-V)} \\ \hline
$[\bar{1}\bar{1}2]$ & $1.0 \times 10^{13}$ & $1.6\times 10^{10}$ & 191 & 3127  &  $6.3 \times 10^{10}$ & $-2.3\times 10^{6}$  \\ \hline
$[1\bar{1}0]$ & $6 \times 10^{13}$ & $1.6 \times 10^{12}$ & 87  & 209  & $1.9 \times 10^{11}$  & $-3.7 \times 10^{10}$ \\ \hline
\end{tabular}
%\end{ruledtabular*}
\end{table*}

  In general, the ratio $n_e/n_h$ is about 50 for the [$1\bar{1}0$] direction, but $\sim10^3$ for the [$\bar{1}\bar{1}2$] direction, suggesting that electrons contribute more to the transport in the [1$\bar{1}$2] direction.  As noted above, we are unable to fit the $V_g$ dependence of $R_H$ with the analogous equation for two electron bands for any sensible range of parameters, so we conclude that both electron and holes must contribute to the electrical transport.  If one considers $R_H$ a direct measure of charge density $n$, $R_H \sim 1/n$, then the anisotropy in $R_H$ is unexpected.  However, in the multiband picture, $R_H$ is a function of the densities of carriers in different bands as well as their mobilities.  Since the measured resistivities (and hence mobilities) along the [$1\bar{1}0$] and [$\bar{1}\bar{1}2$] are clearly different for $V_g \lesssim 20$ V, it is perhaps not surprising that $R_H$ for the Hall bars along two different axes also differs significantly below this gate voltage.  The anomalous gate voltage dependence of $R_H$ that we observe in these (111) samples, as well as the anomalous dependence observed by many groups in the (001) and (110) LAO/STO interface devices, suggests that using $R_H$ to estimate the carrier density $n$ based on a single band model using the equation $R_H = 1/nq$ may not give a correct estimate for $n$.

\subsection{Capacitance Measurements}
\label{cap}
\begin{figure}[t!]
	\includegraphics[width=8cm]{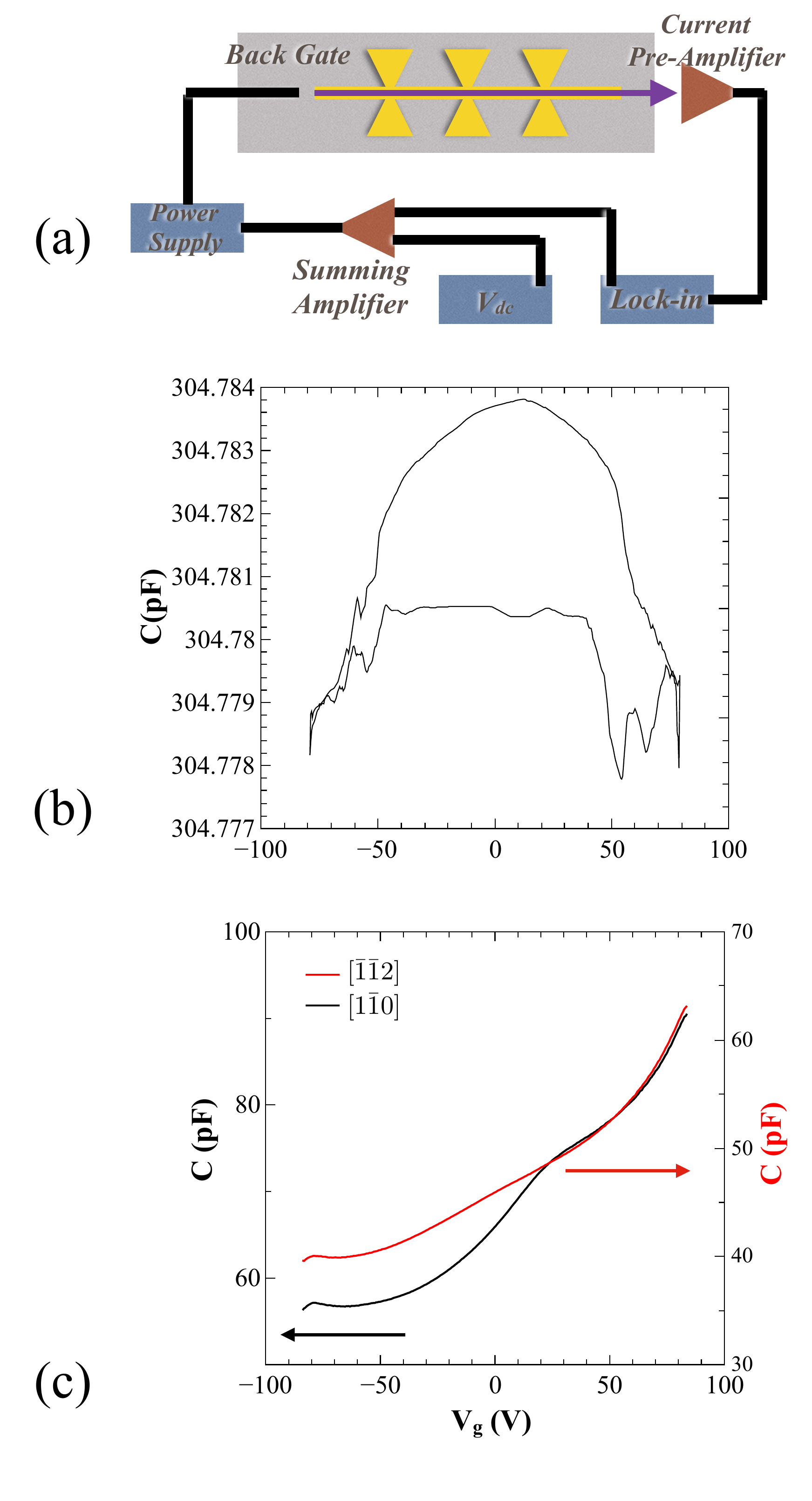}
	\caption{\small   \textbf{a)} Schematic of the capacitance measurement apparatus. \textbf{b)} \small Capacitance of control sample, fabricated on chip with the hall bars, as a function of $V_g$, at 4.4 K. Averaged measured capacitance of the two Hall bars. \textbf{c)} All measurements were performed at 4.4 K.}
	\label{CapRef}
\end{figure}

\begin{figure*}[!]
	\center{\includegraphics[width=15cm]{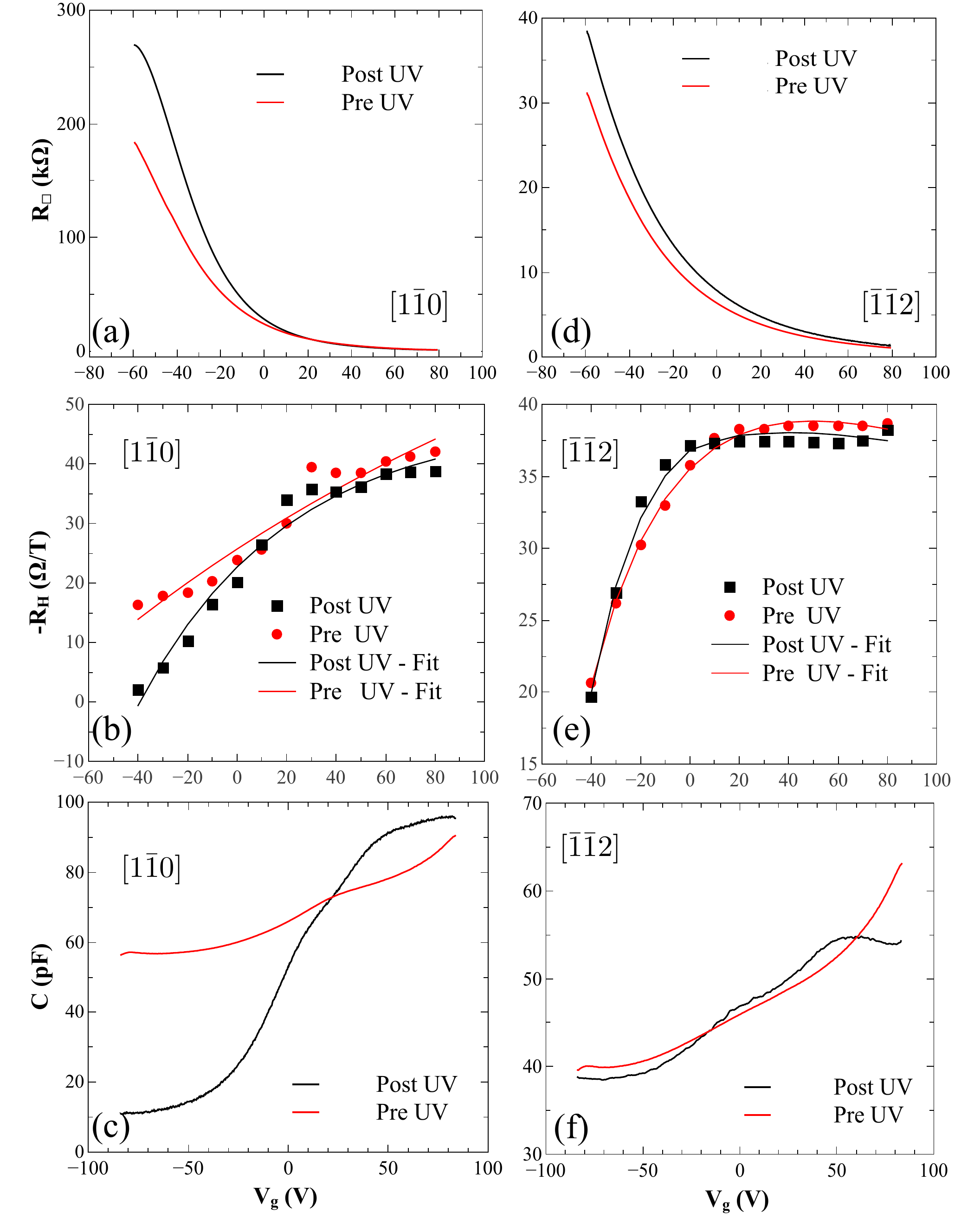}}
	\caption{Comparison of $R_\Box$, $R_H$ and $C$ as a function of $V_g$ before and after UV irradiation.  Red represents data or fits before UV irradiation; black represents data or fits after UV irradiation.  (a), (b) and (c) are for the [$1\bar{1}0$] direction; (d), (e) and (f) are for the [$\bar{1}\bar{1}2$] direction.  Parameters for the fits to Eq. \ref{eq1} after UV irradiation are given in Table \ref{Table2}. All measurements were performed at 4.4 K. }
	\label{Fig3}
\end{figure*}

  For a low-density two-dimensional carrier gas, $C$ is the series combination of the conventional geometric capacitance $C_g$ and the so-called quantum capacitance $C_Q =e^2 S (dn/dE)$,\cite{Eisen,Li2} which is associated with the change in chemical potential resulting from the induced charge.  Here $S$ is the area of the capacitor and $dn/dE$ the density of states.  In our devices, $C_g$ can be determined by measuring the capacitance between the back gate and a large metallic electrode deposited directly on the etched STO substrate of the same chip.  Figure \ref{CapRef}(b) shows $C_g$ measured as a function of $V_g$ measured at 4.4 K.  Normalized to the area, $C_g$ is large, which agrees with the fact that the dielectric constant of STO is large at low temperatures \cite{Li2}. More importantly, it varies by less than 2 parts in 10$^5$ over the full range of $V_g$, so that any variation of $C$ with $V_g$ is related to the quantum capacitance $C_Q$, and hence $dn/dE$.   

Figure \ref{CapRef}(c) shows $C$ for the two Hall bars as a function of $V_g$.  Due to the different effective areas of the Hall bars, comparison of the magnitude of $C$ between two different Hall bars is not meaningful (see Appendix \ref{CapacitanceAppendix}), so here we concentrate on the $V_g$ dependence of $C$.  Overall, the trend for both crystalline directions is that $C$ (and by extension $dn/dE$) is larger at positive $V_g$ in comparison to negative $V_g$, in agreement with the corresponding measurements of $R_\Box$, which decreases with increasing $V_g$ (Fig. \ref{gateh})(a)).    
For $V_g \geq 20$ V, the gate dependence of $C$ in both directions is nearly identical, as can be seen in the figure, where the data have been scaled so that they align in this gate range.  For $V_g \leq 20$ V, however, $C$ for the [$1\bar{1}0$] Hall bar shows a sharp drop in comparison to $C$ for the [$\bar{1}\bar{1}2$] Hall bar.  $C$ for both Hall bars appears to saturate at negative $V_g$; indeed the [$1\bar{1}0$] Hall bar shows hints of an increase in $C$ with decreasing $V_g$ below $V_g \sim -60$ V, a feature that is more prominent on some of the other samples we have measured.  For a purely electron-like $dn/dE$, one would expect to see a monotonic decrease in $C$ with decreasing $V_g$:  the saturation or increasing $C$ that we observe at negative $V_g$ is further evidence that holes participate in electrical transport.

\subsection{Effects of ultraviolet irradiation}
\label{UV}
The data of Fig. \ref{CapRef}(c) show that there is a clear difference in the transport characteristics between the [$1\bar{1}0$] and [$\bar{1}\bar{1}2$] crystal directions that occurs below a gate voltage of $V_g \sim 20$ V.  This is also seen in the Hall data, where there is a sharp drop in $R_H$ in the [$1\bar{1}0$] direction that occurs at the same voltage (Fig. \ref{HallPreUV}), and that is observed in multiple devices:  the feature in $R_H$ is correlated with a corresponding drop in $C$ vs $V_g$ for the [$1\bar{1}0$] direction in comparison to the [$\bar{1}\bar{1}2$] direction; $V_g \sim 20$ V is also roughly the voltage below which the resistances of the two crystal orientations start to diverge strongly. 
Thus, in all three measurements on the (111) interface samples discussed above: $R_\Box$, $R_H$ and $C$, the electrical properties along the 
along the [$1\bar{1}0$] and [$\bar{1}\bar{1}2$] crystal directions are qualitatively and quantitatively similar for $V_g \geq 20$ V, but start differing significantly for $V_g \lesssim 20$ V.  This transition as a function of $V_g$ is sharper in the data for the capacitance and Hall coefficients, and more gradual but larger in the data for the sheet resistance.  At present, the origin for the anisotropy is not clear, but we speculate that it might be due to differences in the carrier bands along the two crystal directions.  Varying $V_g$ changes the chemical potential in the 2D gas at the interface.  When the    chemical potential falls below a band edge, one might expect $dn/dE$ to decrease and $R_\Box$ to increase sharply.  The fact that we observe a sharp drop in $dn/dE$ and a rapid increase in $R_\Box$ for $V_g \lesssim 20$ V measured along the [$1\bar{1}0$] direction, but not along the [$\bar{1}\bar{1}2$] direction leads us to speculate that there are significantly fewer states that contribute to electrical transport for $V_g \lesssim 20$ V along the [$1\bar{1}0$] direction. 
\begin{table*}[!]
\centering
\caption{Post-UV irradiation Hall fit parameters}
\label{Table2}
\begin{tabular}{|l|r|r|r|r|r|r|}
\hline
 & \multicolumn{1}{c|}{$n_e$ ($V_g=0$) (/cm$^2$)} & \multicolumn{1}{c|}{$n_h$ ($V_g=0$) (/cm$^2$)} & \multicolumn{1}{c|}{$\mu_e$ (cm$^2$/V-s)} & \multicolumn{1}{c|}{$\mu_h$ (cm$^2$/V-s)} & \multicolumn{1}{c|}{$dn_e/dV_g$ (/cm$^2$-V)} & \multicolumn{1}{c|}{$dn_h/dV_g$ (/cm$^2$-V)} \\ \hline
$[\bar{1}\bar{1}2]$ & $1.4 \times 10^{13}$ & $1.0\times 10^{10}$ & 121 & 3201  &  $5.3 \times 10^{10}$ & $-1.2\times 10^{6}$  \\ \hline
$[1\bar{1}0]$ & $5.8 \times 10^{13}$ & $2.4 \times 10^{12}$ & 80  & 120 & $1.3 \times 10^{11}$  & $-5.6 \times 10^{10}$ \\ \hline
\end{tabular}

\end{table*}

Band structure calculations and ARPES data from other groups indicate that the bottom of the conduction bands in both directions should be at the same energy, so that one might expect to see similar gradual changes in both directions until the common band edge is reached.  However, these calculations are applicable to higher temperatures:  the band structure calculations are based on room temperature crystal structure, and the lowest ARPES measurements were performed at 20 K.  Our measurements at higher temperatures show little or no anisotropy, and hence are consistent with these band structure calculations and ARPES measurements.  However, we believe our data suggests that, at low temperature, the band edge in the [$1\bar{1}0$] direction is at higher energy than the band edge in the [$\bar{1}\bar{1}2$] direction, and that conduction below the band edge is determined by transport through defect states.

Further evidence of this point of view can be seen in the change in the measured characteristics of the devices ($R_\Box$, $R_H$ and $C$) after irradiation with ultraviolet (UV) light. Historically, irradiation by UV has been conducted in vacuum\cite{Walker}. This irradiation has been shown via transport\cite{Gop}, ARPES\cite{Walker}, and photoluminescence\cite{Kan} to cause an increase in the oxygen vacancies in the sample, which gives
a corresponding increase in the carrier density as each oxygen vacancy donates two electrons to
the interface. The increase in electrons at the interface yields a lower longitudinal
resistance at the interface. In contrast, annealing at elevated temperatures in O$_2$ has been shown to decrease the
oxygen vacancies thereby increasing the resistance at the interface\cite{Liu}.  In the present case, we have irradiated the samples (after making the measurements discussed above) using a UV source placed 1 mm above the samples.  This has the effect of increasing the sample resistance at room temperature, the extent of the increase depending on the exposure time.  Since this is similar in effect to what happens with O$_2$ annealing, we speculate that atmospheric oxygen is introduced into the sample in the form of ozone, reducing the number of oxygen vacancies.  The effect is transient at room temperature:  the sample regains its original resistance over a period of minutes to hours, depending on the initial exposure time.   
However, the induced change persists if the sample is immediately cooled down to liquid nitrogen temperatures.  For the current study, we irradiated the sample for 15 minutes under these conditions, which increased the room temperature sheet resistance from 22.15 k$\Omega$ to 33.66 k$\Omega$.  After exposure, the samples were cooled down as soon as possible to liquid nitrogen temperatures for measurement.

This treatment appears to disproportionately affect transport in the [$1\bar{1}0$] direction for $V_g \lesssim 20$ V.  Figure \ref{Fig3} shows a comparison of the data for the two crystal directions before and after 15 minutes of UV irradiation.  (The data for before UV irradiation are the same as that shown in Figs. \ref{gateh}(a), \ref{HallPreUV} and \ref{CapRef}(c).  For $R_\Box$, the resistance in both directions at negative $V_g$ after UV irradiation increases.  In the [$1\bar{1}0$] direction, the pre- and post-UV sheet resistances  start to diverge significantly below $V_g\sim 20$ V, while the difference in the pre/post UV $R_\Box$ for the [$\bar{1}\bar{1}2$] direction grows steadily over the entire $V_g$ range.  More notable differences can be seen in $R_H$ and $C$.  While $R_H$ for the [$\bar{1}\bar{1}2$] direction shows little change after UV treatment, $R_H$ for the [$1\bar{1}0$] direction starts to decrease in comparison with the pre-UV values for $V_g\lesssim 20$ V:  at $V_g=-40$ V, $R_H$ is close to 0.  However, the changes in $C$ are the most striking.  For the [$\bar{1}\bar{1}2$] direction, $C$ before and after UV exposure is approximately the same, except for a small range of $V_g$ above 65 V.  For the [$1\bar{1}0$] direction, however, $C$ drops off sharply as one goes from positive to negative $V_g$, indicating a corresponding drop in $dn/dE$ at around $V_g \sim 20$ V, the same voltage range at which the differences between pre- and post-UV in the measured $R_H$ and $R_\Box$ become apparent.  (Note that here, comparison of the magnitudes of $C$ are meaningful, as the data are for the same Hall bar, and $S$ has not changed.)  These data strongly suggest that the number of electrons available to participate in transport at $V_g \lesssim 20$ V for the [$1\bar{1}0$] direction decreases substantially after UV irradiation. This strengthens our speculation that we reach the band edge at higher $V_g$ for the [$1\bar{1}0$] direction in comparison to the [$\bar{1}\bar{1}2$] direction; the differences in pre- and post-UV measurements suggest further that the states that contribute to conduction below the band edge are reduced by the UV treatment.

\section{Summary}
We have shown a clear and surprising anisotropy at the (111) LAO/STO interface between Hall bars fabricated along the orthogonal [$1\bar{1}0$] and [$\bar{1}\bar{1}2$] directions. Specifically, we have found that the this anisotropy can be tuned via the application of $V_g$. As $V_g$ is tuned to high positive voltage, pushing carriers to the interface, both directions have similar longitudinal resistances, Hall characteristics, and capacitances. However, when $V_g$ is tuned to large negative voltages, the longitudinal resistance, Hall coefficients and capacitance differ significantly along the two mutually orthogonal crystal directions.  This anisotropy can be made much stronger by irradiating the sample with UV light, but interestingly, the voltage range over which the anisotropy becomes significant ($V_g \leq 20$V) stays the same. 
Our observations are reproduced across multiple sample chips grown in different growth runs, with different substrate preparations and on different cooldowns, showing that the anisotropy is not due to structural changes associated with growth conditions, but is an intrinsic property of the (111) LAO/STO interface.  The fact that we do not observe anisotropy at 77 K, below the STO structural transition temperature of 105 K, also indicates that the anisotropy observed at liquid helium temperatures and below is also not solely associated with this structural transition, although the lowering of the crystal symmetry associated with the transition may play a role.

As we noted above, the data suggest that the anisotropy arises from a difference in the conduction band edge along the two crystalline directions.  Other groups have reported splitting of the bands at the band minima due to spin-orbit interactions in (001) \cite{Santander} and (110) \cite{Gop} LAO/STO heterostructures.  On the other hand, band structure calculations\cite{Walker} and ARPES\cite{Rodel} data on (111) heterostructures from other groups indicate that the bottom of the conduction bands in both directions should be at the same energy, so that one might expect to see similar gradual changes in both directions until the common band edge is reached.  However, these calculations and measurements were for samples at room temperature or 20K: our data indicate that there might be a change in the band structure at liquid helium temperatures that accounts for our observations.  Thus, it is quite likely that a band splitting might develop at lower temperatures in the(111) LAO/STO heterostructures as well. 

In addition, we have observed evidence for hole like carrier contributions to the transport properties in Hall bars along both crystal directions at the (111) interface. The strongest evidence for this contribution comes from the dependence of the Hall coefficient, $R_H$, as a function of $V_g$: $R_H$ increases over the measured voltage range of -40 to 80V, and this behavior cannot be explained solely by a multiple electron model. This dependence can only be explained if there is a contribution from at least one hole band, which may arise from defect acceptor levels just above the valence band edge that have been described in earlier work.\cite{Kan}

\begin{acknowledgments}
We would like to thank Varada Bal for performing the initial photo-lithography steps, and Andy Millis for useful discussions.  Work at Northwestern was funded through a grant from the U.S. Department of Energy through Grant No. DEFG02-06ER46346.
Work at NUS was supported by the MOE Tier 1 (Grant No. R-144-000-364-112 and R-144-000-346-112) and Singapore National Research Foundation (NRF) under the Competitive Research Programs (CRP Award No. NRF-CRP8-2011-06 and CRP Award No. NRF-CRP10-2012-02).

\end{acknowledgments}

%Appendices
\appendix

\section{X-ray characterization}
To determine the crystal orientation of the Hall bars shown in Fig. 1(b) of the manuscript, we utilized a Photonic Science Laue X-ray camera. The sample was mounted on a goniometer stage for precise alignment. The images were fit using the commercial PSL software provided with the camera, which allowed simultaneous determination of both Hall bar orientations on the substrate, which are as marked in Fig. 1(b) of the main manuscript.

\section{Capacitance measurement and analysis}
\label{CapacitanceAppendix}
The capacitance that we measure between the back gate and a Hall bar sample includes the entire area of the Hall bar, which in addition to the Hall bar itself, also includes the voltage and current contacts on which we make the wire bonds.  In fact, the area of the contacts is a substantial fraction of the total area of the device.  However, the wire bonds punch through to the interface gas, decreasing the area of the capacitor, and since the each wire bond is different, the effective area of each Hall bar capacitor is different.  Consequently, we cannot compare the absolute values of the measured capacitance of two different Hall bars.  However, for the same Hall bar before and after UV irradiation, a comparison is meaningful, since we do not remove the wire-bonds during the irradiation process.

A capacitor with some leakage can be described as a pure capacitance $C$ in parallel with a low conductance $G$.  The admittance $Y$ of this parallel arrangement is then simply $Y=G + j\omega C$, where we have employed engineering notation, so that the quadrature signal is a direct measure of the capacitance.

\begin{figure}[h!]
\includegraphics[width=8cm]{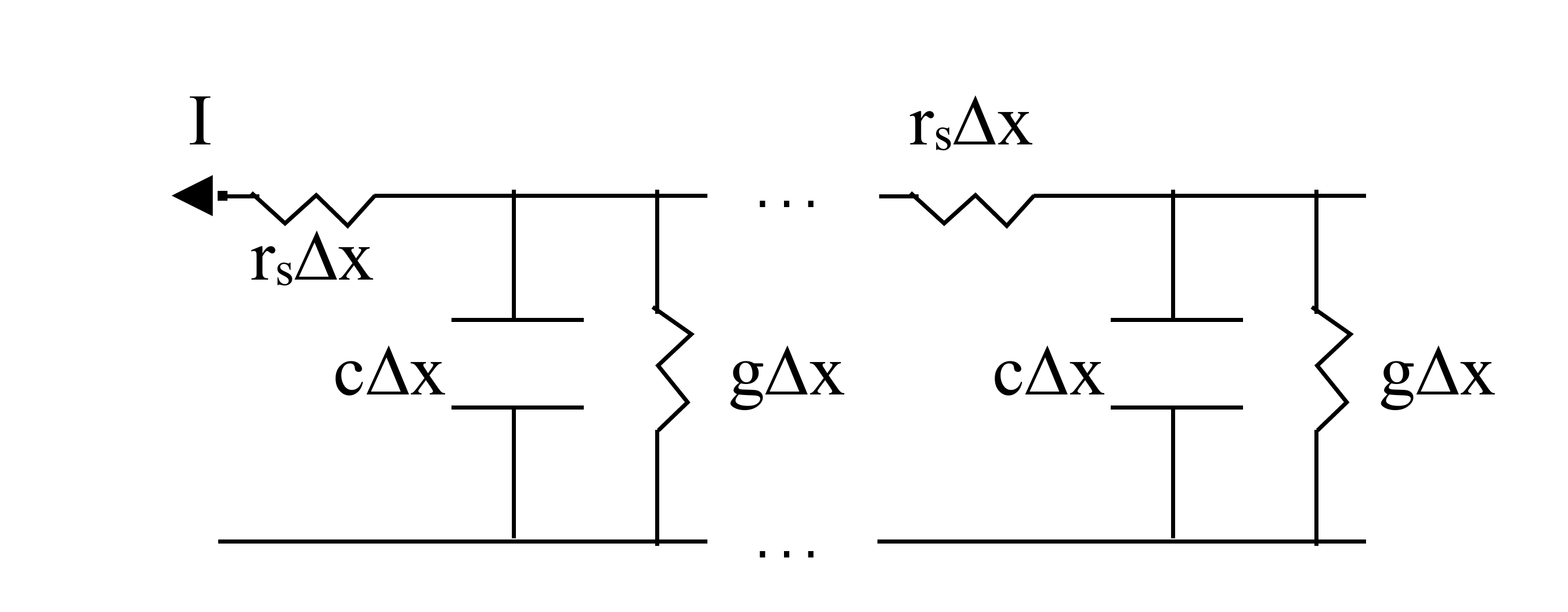}
\caption{\small   Model diagram of the measurement $R_S$ corresponds to the sheet resistance of the 2DEG, $C_T$ corresponds to measured capacitance, and $G$ the conductance of the STO substrate. }
\label{CapSchem}
\end{figure}   
\begin{figure}[h!]
\includegraphics[width=7cm]{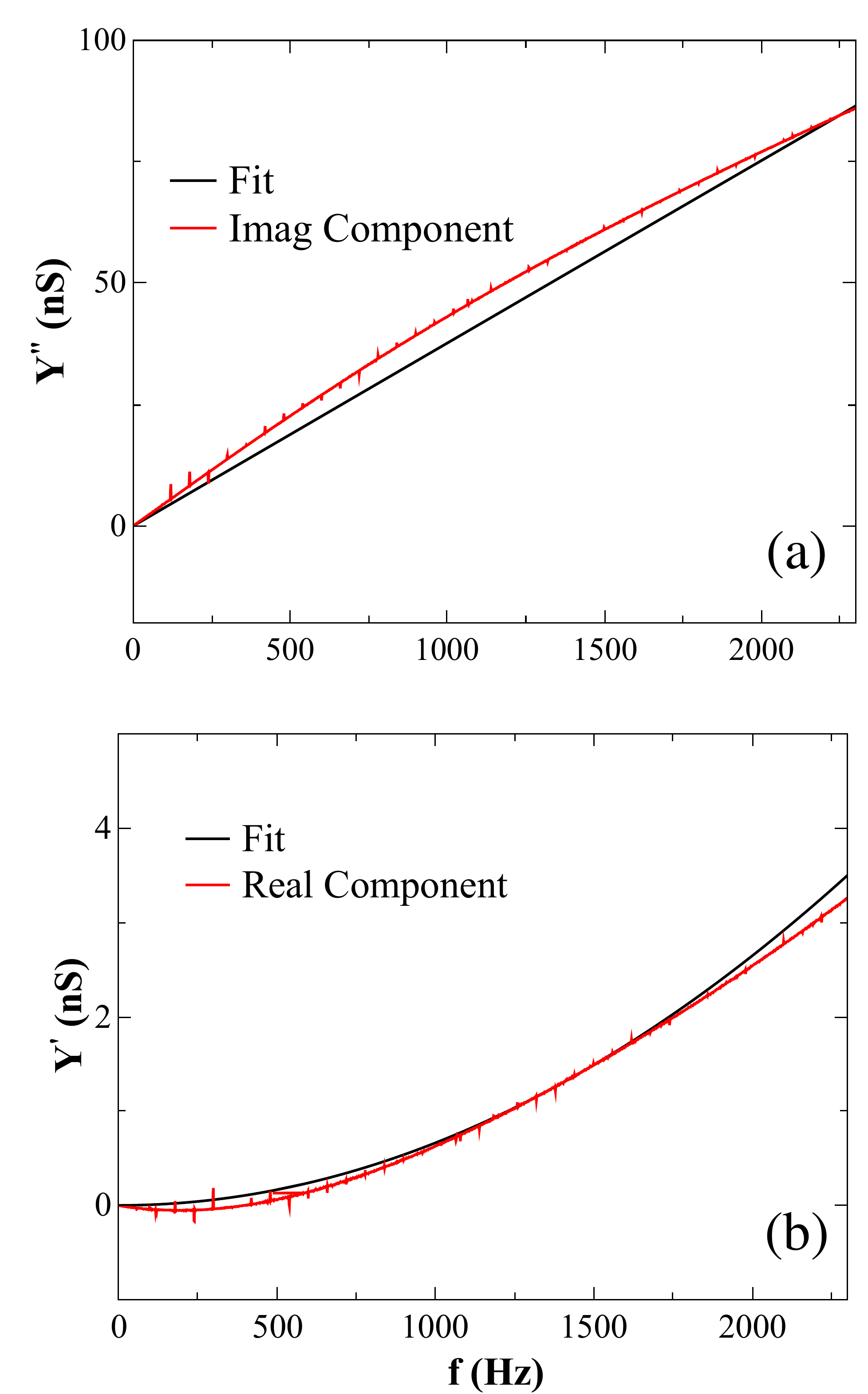}
\caption{\small Real and imaginary components of the admittance as a function of the frequency $f=\omega/2\pi$ of the $[1\bar{1}0]$ Hall bar at 4.4 K, measured as shown in Fig. \ref{CapSchem}(a) for $V_g = X$ V.   The black curves are fits to Eq. S\ref{TransLine} with parameters $R_s= 25.4k\Omega /\square$, $G=1E-15S$ and $C=63.5pF$.}
\label{FreqDepFit}
\end{figure}  
However, the conducting interface can also have appreciable resistance, depending on $V_g$, and in order to determine whether this resistance might account for the anisotropy in the measured capacitance that we observe, we have modeled the sample as a transmission line with a capacitance per unit length $c=C/L$, a conductance per unit length $g=G/L$, and an interface resistance per unit length $r_s=R_s/L$, as shown in Fig. \ref{CapSchem} (here $L$ is the total length of the sample).  The transmission line analysis leads to the admittance
\begin{align}
Y &= \frac{\sqrt{r_s(g+j\omega c)}}{r_s} \tanh(\sqrt{r_s(g+j\omega c)} L) \nonumber \\
&= \frac{\sqrt{R_s(G+j\omega C)}}{R_s} \tanh\sqrt{R_s(G+j\omega C)}
\label{TransLine}
\end{align}

\begin{figure}[h!]
\includegraphics[width=7cm]{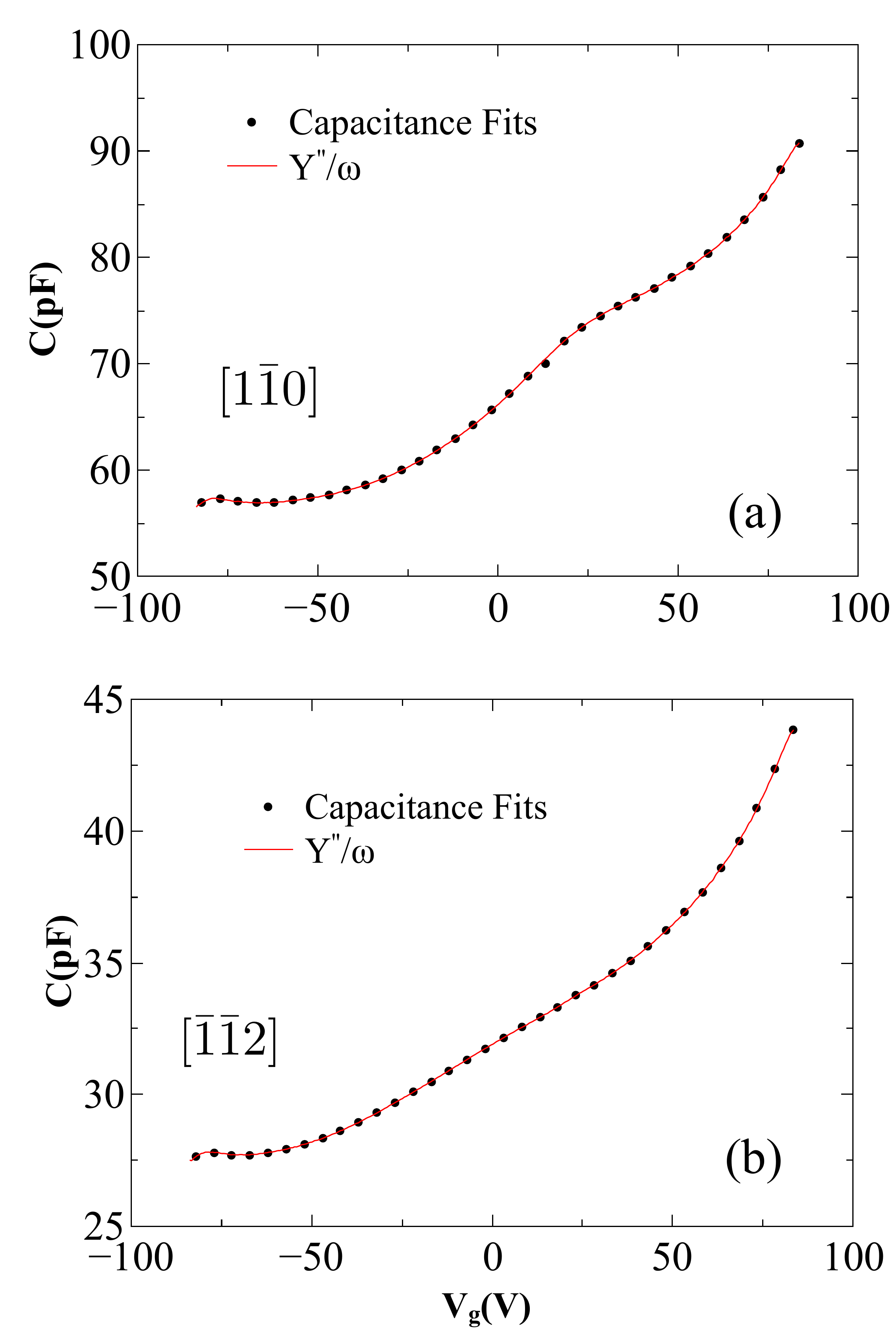}
\caption{Solid symbols:  Capacitance obtained by simultaneously fitting the real and imaginary components of the measured admittance to Eq. \ref{G0limit} at each value of $V_g$, for both Hall bars at 4.4K.  Solid lines are the measured quadrature component divided by $\omega$.}
\label{YImagFits}
\end{figure} 

Figure \ref{FreqDepFit} shows the frequency dependence of the real and imaginary components of $Y$ measured for the $[1\bar{1}0]$ Hall bar for $V_g= 0V$ at 4.4 K.  The fits are quite reasonable; we believe that the small discrepancy between the fits and data is due to the phase dependence of the Kepco, since the lock-in amplifier is phased at one fixed frequency before the frequency sweep.  The values of $R_s$ and $G$ are also reasonable:  $R_s$ agrees to within 10 \% of the measured resistance of the sample at that value of $V_g$, which we consider good agreement.  The values obtained for $G$ are $\sim 10^{-15}$ S.  In fact, the fits are insensitive to $G$ in this range, and we take $G=0$ in our further analysis.  With $G=0$, Eq. \ref{TransLine} gives the following expressions for the real and imaginary components of $Y$ ($Y=Y' + jY''$)
\begin{align}
Y' &= \frac{\sqrt{2 \omega C R_s}}{2R_s} \frac{\sinh \sqrt{2 \omega C R_s} - \sin \sqrt{2 \omega C R_s}}{\cosh \sqrt{2 \omega C R_s} + \cos \sqrt{2 \omega C R_s}}, \nonumber \\
Y'' &=  \frac{\sqrt{2 \omega C R_s}}{2R_s} \frac{\sinh \sqrt{2 \omega C R_s} + \sin \sqrt{2 \omega C R_s}}{\cosh \sqrt{2 \omega C R_s} + \cos \sqrt{2 \omega C R_s}}.
\label{G0limit}
\end{align}

We have used Eq. \ref{G0limit} to fit the real and imaginary parts of the measured admittance as a function of gate voltage.  The solid symbols in Fig. \ref{YImagFits} show the capacitance $C$ obtained from simultaneously fitting the real and imaginary parts of $Y$ to Eq. \ref{G0limit}.  For comparison, we show the measured imaginary component $Y''/\omega$.  The curves for both samples are identical, showing that the resistance of the interface $R_s$ does not affect the measured capacitance, and that the imaginary component of the admittance is simply $Y''=\omega C$.  In fact, this is the limit of Eq. \ref{G0limit} for $\omega R_s C << 1$.  Consequently, we can determine the capacitance directly from the quadrature signal.  It should be noted that the insensitivity of $Y''$ to $R_s$ in our parameter range means that the anisotropy that we observe in the measured capacitance is not due to the anisotropy in the sheet resistance of the two Hall bars.

\end{document}